\documentclass[apjl]{emulateapj}
\newcommand{\teff}{$T_{\rm eff}$}

\shorttitle{LITHIUM AND PROTON-CAPTURE ELEMENTS IN GC DWARFS}
\shortauthors{D'ORAZI ET AL.}

\begin{document}


\title{LITHIUM AND PROTON-CAPTURE ELEMENTS IN GLOBULAR CLUSTER DWARFS:  \\
THE CASE OF 47 TUC}


\author{Valentina D'Orazi\altaffilmark{1}}
\author{Sara Lucatello\altaffilmark{1,2}}
\author{Raffaele Gratton\altaffilmark{1}}
\author{Angela Bragaglia\altaffilmark{3}}
\author{Eugenio Carretta\altaffilmark{3}} 
\author{Zhixia Shen\altaffilmark{4}}
\author{Simone Zaggia\altaffilmark{1}}
\altaffiltext{1}{INAF--Osservatorio Astronomico di Padova, 
vicolo dell'Osservatorio 5 , I-35122, Padova, Italy}
\altaffiltext{2}{Excellence Cluster Universe, Technische Universit\"{a}t 
M\"{u}nchen,
Boltzmann Str. 2, D-85748, Garching, Germany}
\altaffiltext{3}{INAF--Osservatorio Astronomico di Bologna, via Ranzani 1, 
I-40127, Bologna, Italy}
\altaffiltext{4}{National Astronomical Observatories, Chinese Academy of
Science, 20a Datun road, Beijing, China}
\email{valentina.dorazi@oapd.inaf.it}

\begin{abstract}

Previous surveys in a few metal-poor globular clusters (GCs) showed that the determination of 
abundances for Li and proton-capture 
elements offers a key tool to address the intracluster pollution scenario. In this Letter, 
we present Na, O, and Li 
abundances in a large sample of dwarf stars in the metal-rich GC 47 Tucanae. 
We found a clear Na$-$O anticorrelation, in good agreement with what 
obtained for giant members by Carretta et al. While lithium and oxygen abundances appear 
to be positively correlated with each other, there is a large scatter, well exceeding observational errors, and no anticorrelation with sodium. 
These findings suggest that Li depletion, due to mechanisms internal to the stars 
(which are cooler and more metal-rich than those on the Spite plateau) combines with 
the usual pollution scenario, responsible for the Na$-$O anticorrelation.
\end{abstract}

\keywords{globular clusters: individual (NGC~104) -- Stars: abundances -- 
stars: individual (Population II)}

\section{Introduction}\label{sec:intro}

The traditional paradigm of globular clusters (GCs) as the textbook example 
of a simple stellar population is now
outdated. In the last several years, a wealth of observational 
studies (both photometric and spectroscopic) have been carried out, 
revealing the quite complex nature of this old Galactic population. 

With the famous exception of $\omega$~Cen (Freeman \& Rodgers 1975), and of the more recently scrutinized
M~54 (Carretta et al. 2010), M22 (Marino et al. 2009), and Terzan 5 
(Ferraro et al. 2009), GCs show homogeneous compositions in the iron peak 
and heavier $\alpha$ elements (e.g., Ca and Ti). 
Abundance variations in the lighter elements, namely Li, C, N, O, Na, Mg 
and Al, have been instead recognized since the 1970's 
(e.g. Cohen 1978; see Gratton, Sneden \& Carretta 2004 for a recent review). 
These peculiarities of GC stars require that some material must have been processed through the 
complete CNO cycle in hot H burning, through proton-capture reactions (Denisenkov \& Denisenkova 1990): element pairs C and N, O and Na, and Mg and Al are anticorrelated, the 
abundances of C, O and Mg being depleted while those of N, Na and Al being enhanced.

A previous generation of stars, which synthesized proton-capture elements in their interiors, 
is now commonly accepted as responsible for the above-mentioned chemical signatures. 
Regardless of the debated nature of these element polluters (asymptotic giant
branch (AGB) stars experiencing hot bottom burning, e.g., Ventura \& D'Antona 2009, 
or fast rotating massive stars (FRMS), e.g., Decressin et al. 2007), two fundamental observational facts deserve to be emphasized. 
(1) All the GCs surveyed so far show the Na$-$O anticorrelation (Carretta et al. 2009b): this indicates the presence of at 
least two populations (neither coeval nor chemically homogeneous) within each cluster. 
(2) The primordial nature of such a phenomenon is uniquely indicated by the 
presence of these chemical signatures also among not evolved turnoff (TO) or scarcely evolved (subgiant, SGB) members 
(see, e.g., Gratton et al. 2001).  

In this context, Li abundances play a fundamental role. In fact, this element can be easily destroyed in stellar interiors 
(starting at T$_{\rm burn}$$\approx$2.5$\times$10$^6$~K); and since the CNO and NeNa cycles require much higher temperatures, 
it is expected  that in the site(s) where these cycles occur no Li is left. In particular, the 
Na-poor, O and Li rich stars, that are the first population born in the cluster, share the chemical composition of field stars of the same metallicity, 
while the Na-rich (Li- and O-poor) stars form from gas progressively enriched by ejecta (which are rich in Na, depleted in O and Li) of the first generation. 
 
As a consequence, if primordial and processed material are mixed in different 
proportions, then Li and Na (Li and O) are expected to be anticorrelated (correlated) with each other.

Measuring the Li abundances in unevolved stars provides a direct indication of the amount of pristine, and by difference of polluted (CNO-cycle processed), material present in each star. This makes Li a unique tracer of the dilution process which took place in the GC, supplying also fundamental insights on its early evolutionary stages. 
Another important point is that 
Li abundances can provide strong observational constraints on the origin of the polluters: since AGB stars might also produce 
Li (via the Cameron$-$Fowler mechanism, Cameron \& Fowler 1971), while massive stars can only destroy it, if Li-rich stars 
are present in GCs, AGB stars would be definitely favored with respect to FRMS.
 
To date, only three GCs have been surveyed for correlations between Li and proton-capture elements. 
Pasquini et al. (2005) obtained Li abundances in nine TO members of NGC 6752: they found a depletion reaching down to
$\sim$1 dex below the Spite plateau values, clearly anticorrelated with Na abundances. 
A similar result was obtained by 
Bonifacio et al. (2007) who found a scatter in Li abundances much larger than observational errors and 
anticorrelated with Na among (only) four stars in 47 Tuc. Very recently, Lind et al. (2009), targeting about 100 main-sequence (MS) and 
early subgiant branch  
stars in NGC~6397, detected for the first time a significant anticorrelation between Li and Na in this GC. 
This last investigation supersedes an older research by Bonifacio et al. (2002) based on only a few stars, and 
highlighting the importance of large samples of stars for similar studies.

In this Letter we present Na, O, and Li abundance determinations in $\sim$90 unevolved TO stars of 
the GC 47 Tucanae, 
providing the {\it largest} database of this kind available in the literature so far. 

\section{Sample and Analysis}\label{sec:analysis}

We retrieved from the ESO Archive 
FLAMES-Giraffe spectra (MEDUSA configuration) of 109 unevolved members of the metal-rich GC 47 Tuc (Program 081.D-0287; PI Z. Shen). 
The observations were carried out in 
Service mode in 2008 August--September; gratings HR15n ($R\sim$17,000), HR18 ($R\sim$18,400), and 
HR20A ($R\sim$16,000) were used to cover Li~{\sc i} (6707 \AA\*), 
O~{\sc i} (7771, 7773, 7774 \AA\*), and Na~{\sc i} (8183, 8194 \AA\*) features, respectively.

The data reduction was performed with the ESO Giraffe 
pipeline\footnote{\url{http://www.eso.org/sci/data-processing/software/pipelines\\
/giraffe/giraf-pipe-recipes.html}} using {\sc Gasgano} and
following the standard procedure: bias subtraction, flat-field correction, 
wavelength calibration, and optimal spectrum extraction. 
A careful sky subtraction has been done using the IRAF\footnote{IRAF is distributed by the National Optical Astronomy Observatory, 
which is oprated by the Association of Universities for Research in Astronomy, Inc., under cooperative agreement
with the National Science Foundation.} task {\it skytweak} 
in order to take into account wavelength shifts and possible differences 
in emission line strengths between stellar and sky spectra. It is noteworthy that 
the presence of the moon at more than 90\% illumination along with 
diffuse clouds caused some of the observed frames to be severely contaminated 
and were thus discarded (the night of August 17th). Different exposures were 
then co-added to improve signal-to-noise ratios, obtaining typical values of 
$\sim$50--100 per pixel (near the Li~{\sc i} region at $\sim$6708 \AA\*). 
  
The abundance analysis was carried out using the {\it ROSA} code (developed by R. Gratton) and Kurucz (1993) model atmospheres; 
for Na and O, we derived abundances from the equivalent widths (EWs), applying corrections due to non-LTE effects (Gratton et al. 1999). 
LTE Li abundances were instead obtained through spectral synthesis. Stellar parameters were derived as follows. (1) We obtained initial 
effective temperatures (\teff)
from the de-reddened ($E(B-V)$=0.04; Harris 1996) $B-V$ 
colors\footnote{Data are from unpublished photometry by Y. Momany (private communication);  
they are derived and calibrated exactly as the other dat sets in Momany et al. (2003).} 
and the calibration by Alonso et al. (1996). The adopted
values were derived from a relation between $V$ magnitude and color-based \teff. This reduces errors for individual stars. (2) 
The surface gravities were estimated from the fundamental relation between $M$, $L$, and \teff~
($\log g=4.44+ \log(M/M_\odot)-\log(L/L_\odot)+
4\log T_{\rm eff} - 15.0447$), by adopting the bolometric corrections by Alonso et al. along with a distance 
modulus of 13.34 (Harris 1996), while $M_{\rm TO}$ was calculated in the same fashion described in Gratton et al. (2010).  

Finally we assumed for all our sample stars microturbulence $\xi$=1.00 km s$^{-1}$ 
(note that none of the elements under scrutiny is strongly affected by $\xi$ values).

The [X/Fe] ratios were calculated assuming as solar references $\log~n{\rm (Na)}_{\odot}$=6.21 and $\log~n{\rm (O)}_{\odot}$=8.79, values obtained in a fully consistent fashion from the analysis of 
the solar spectrum by Gratton et al. (2003), and [Fe/H]=$-$0.76 (Carretta et al. 2009a) as cluster metallicity.

Both internal (random) and external (systematic) 
errors may affect our analysis. A discussion of the first kind of uncertainty, 
which is predominant for investigating star-to-star variations, can be found 
in Carretta et al. (2007); here we just mention that errors in EWs are dominant,
while the internal uncertainties on stellar parameters (\teff)
can be neglected (note that the distance of a star on the color-magnitude diagram from 
the MS main loci represents the major error source in the adopted temperature values). 
Systematic errors due to the adopted set of model atmospheres 
and/or to reddening and T$_{\rm eff}$ scale cannot be larger than $\sim$0.1 dex (see Bonifacio et al. 2002, 2007 and Gratton et 
al. 2001); 
in particular, for Li, the combination of three-dimensional and non-LTE effects yields negligible 
corrections (see, e.g., Asplund et al. 2003).

All the abundance values with the corresponding uncertainties, along with magnitudes and coordinates, are available 
only in electronic version.

\section{Results and Discussion}\label{sec:disc}

\begin{figure}
\begin{center}
\includegraphics[width= 8cm]{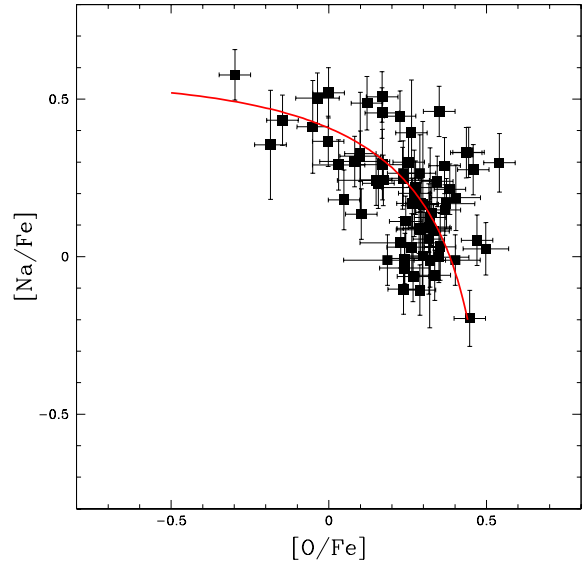}
\caption{Na$-$O anticorrelation for 47 Tuc dwarf members; error bars come 
from uncertainties on EWs. A dilution model (Prantzos \& Charbonnel 2006) is 
also overplotted as a (red) solid line.}\label{f:nao}
\end{center}
\end{figure}

As shown in Figure~\ref{f:nao}, the unevolved cluster stars reveal a very clear Na$-$O 
anticorrelation, consistently with what has been obtained for giant members  
(Carretta et al. 2009b). We detected an oxygen variation of about 
$\sim$0.80\,dex and a change in Na of $\sim$0.77\,dex, to be compared with 
$\sim$1 and 0.82 (O and Na respectively) for giants in 
Carretta et al. (2009b). The interquartile ranges 
IQR[Na/O] are 0.368$\pm$0.067 and 0.470$\pm$0.041, respectively. 
These two estimates are consistent at 1$\sigma$ level; however since the 
dwarfs are a more homogeneous sample (i.e., very similar atmospheric parameters) 
and the spectral features used to measure a given element are not 
the same in dwarfs and giants, the latter being weaker and plagued by higher uncertainties,
 both the spread (maximum--minimum) and the IQR result slightly higher for giants. 
It is noteworthy that we found an offset in Na abundances between dwarfs 
and giants of $\sim$0.2 dex for 47 Tuc. A similar offset can
be obtained also for NGC~6397, when comparing Na abundances for 
dwarfs from Lind et al. (2009) and for giants from Carretta et al. (2009b). 
We investigated the nature of this offset, concluding that it is due 
to the adopted set of lines, with the doublets 5682$-$
5688~\AA\ and 6154$-$6160~\AA\ (used in Carretta et al.) yielding always 
higher abundances than the 8183-8194 \AA\ features (employed for dwarfs). 
We, in fact, detected the same effect in the Sun 
(with the same $\Delta$(Na)$\sim$~0.2 dex). 
We propose that the way damping wings are treated is responsible for the observed difference. 
The present analysis uses values consistent with the 
Barklem et al. (2000) values; however, EWs are extracted using Gaussian
profiles, which may result in systematic underestimates of the EWs for very strong lines, 
like the 8183-8194 \AA~ones. We then consider abundances from the weaker 
doublets as more reliable, and we offset our Na abundances for this effect.
 
\begin{figure}
\begin{center}
\includegraphics[width=8cm]{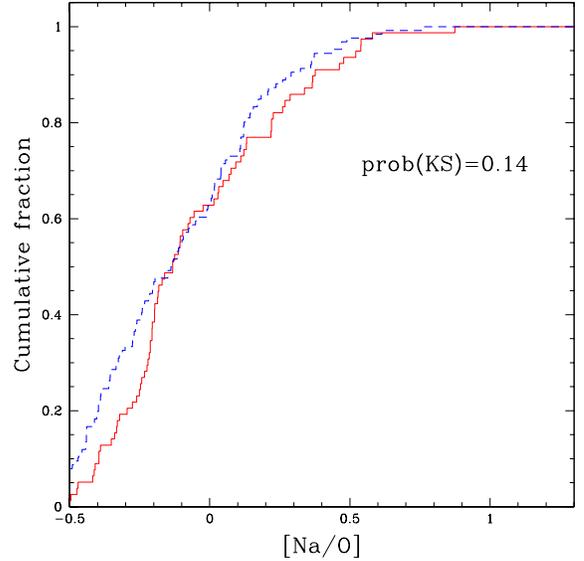}
\caption{Cumulative [Na/O] distribution  for dwarfs (red solid line) and giants (blue
dashed line).}\label{f:ks1}
\end{center}
\end{figure}

In general, we conclude that there is a very good agreement between the two distributions as can be clearly seen in Figure~\ref{f:ks1}: 
after applying the derived offset in Na abundances, the cumulative 
[Na/O] functions look very similar, and a 
Kolmogorov$-$Smirnov (KS) test indicates that 
the two distributions cannot be distinguished 
(the probability from KS being $\sim$14\%).
 
Thanks to the wide sample available, we confirm that evolutionary effects, 
acting during the red giant branch (RGB) phase (see, e.g., D'Antona \& Ventura 2007
for M~13), cannot contribute to the Na$-$O distribution (at least for the present cluster), since the 
extent of anticorrelation in both giants and dwarfs is essentially the same. Furthermore, following the approach by Carretta et 
al. (2009b) the fractions of primordial (P)\footnote{The P stars must have [Na/Fe]$\leq$[Na/Fe]$_{\rm min}$ + 0.3, then the remaining stars are divided in I and E according to [O/Na] larger and smaller 
than $-$0.9 dex, respectively.} , 
intermediate (I) and extreme (E) stars computed from unevolved members are 
34\%$\pm$5\%, 63\%$\pm$7\%, and 3\%$\pm$1\% respectively to be compared with 
27\%$\pm$5\%, 69\%$\pm$8\%, and 4\%$\pm$2\% from giants (see Carretta et al. 2009b). 

\begin{figure*}
\begin{center}
\includegraphics[bb=30 400 600 700,clip,width=15cm]{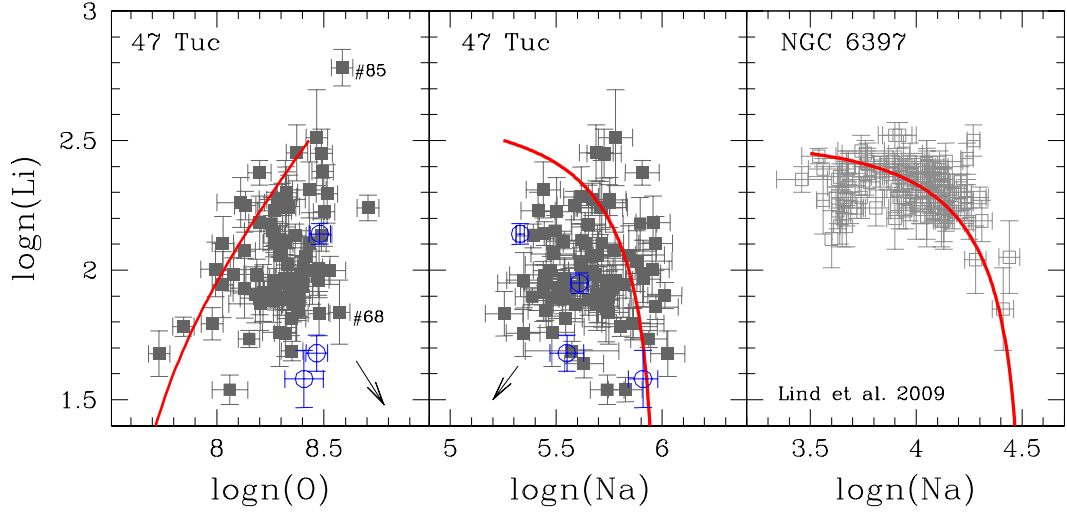}
\caption{Run of Li with O and Na for 47 Tuc (left and middle panels). 
The (blue) open symbols are stars 
with Li from Bonifacio et al. (2007) and Na, O
from Carretta et al. (2004). A dilution model is overplotted 
as (red) solid
lines (see the text).
The arrow represents the effect on abundances of a
change in T$_{\rm eff}$ of 145 K, which is the offset between the two different
temperature scales employed. In the right-hand panel we show, for comparison, the Li$-$Na diagram for the 
GC NGC~6397 from Lind et al. (2009).}\label{f:6397}
\end{center}
\end{figure*}

\begin{figure*}
\begin{center}
\includegraphics[bb=40 180 600 500, clip,width=13cm]{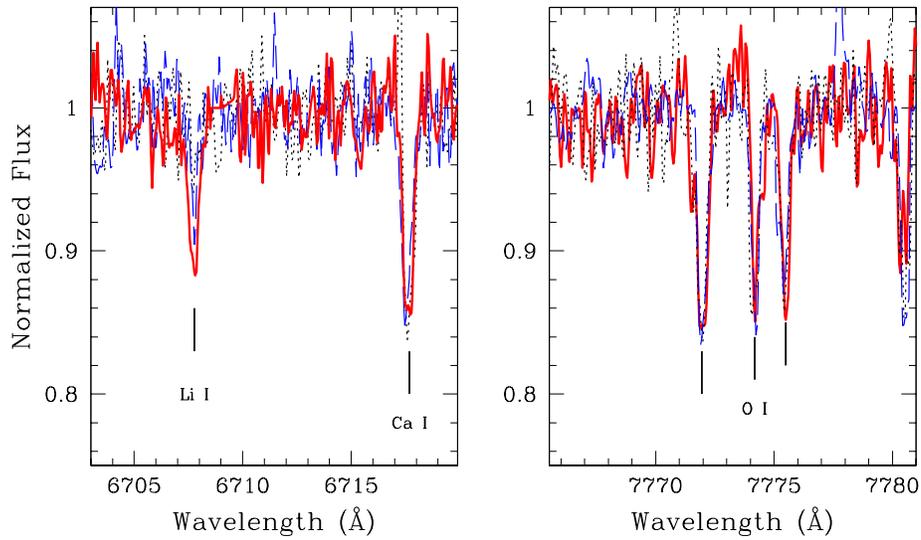}
\caption{Two portions of the spectra, centered on Li (left panel) and O (right
panel), are shown for the stars \#068, \#083, and \#085.}\label{f:spectra}
\end{center}
\end{figure*}

On the other hand, there is not a one-to-one correlation between Li and O abundances 
(left-hand panel of Fig.~\ref{f:6397}): even if
some trend might be present, there is a considerable scatter, much larger than observational errors. The figure shows that, while
the stars with low O values have also low Li content, at higher O abundances, Li 
can assume a large range values, from 
1.54$\pm$0.06 to 2.51$\pm$0.18, with the most extreme star (\# 085) having 
$\log~n{\rm (Li)}=2.78$$\pm$0.08 (this very Li-rich star, which appears as
primordial type according to its O abundances, surely deserves special
attention and further
high-resolution observations, in order to investigate this extreme behaviour). 
In Figure~\ref{f:spectra} we show, as explicative example, the spectra of three 
stars
with almost the same O abundances; one is star \# 085 and the two others have intermediate and low Li.\\
If we then focus on the Li$-$Na
diagram (middle panel of Figure~\ref{f:6397}), 
the scatter is even larger with {\it no evidence for an anticorrelation}. 
As further evidence, the linear correlation coefficient is 0.39 (77 stars) 
and $-$0.02 (84 stars), respectively, for the Li$-$O and Li$-$Na distributions: in the
first case, the significance level is $>$99.5\%, while the second one
has no statistical 
meaning.

Our study increases the sample of Li determination in this cluster, 
completing the previous studies by Pasquini \& Molaro (1997, Li for two stars) 
and 
by Bonifacio et al. (2007), who proposed a Li$-$Na anticorrelation 
on the basis of only four members.

To get more insight into this issue, we can compare it with 
a simple model. It is known that the Na$-$O 
anticorrelation can be well described by a simple
dilution model\footnote{The basic equation of this model is:
$[X]=log[(1-dil)\times10^{\rm[X_O]}+dil\times10^{\rm[X_p]}]$, where $0 < dil < 1$, $[X_{\rm O}]$ and $[X_{\rm
p}]$ are the logarithmic abundances of the element in the original and processed material, respectively.}, like the one proposed
by Prantzos \& Charbonnel (2006). On the other hand, 
Li abundances show a wide scatter, with values often well
below those expected from a simple dilution model 
(red solid line in three panels of in Figure~\ref{f:6397}). 
However, we cannot exclude that this model might represent 
the upper envelope of the Li distribution. 
{\it In the case of 47 Tuc, Li is telling us something 
different from Na and O. Different mechanism(s) have to be 
invoked in order to explain the observed abundance pattern.}

The distribution of Li abundances for 47 Tuc hence looks very different from that found for the 
metal-poor ([Fe/H]=$-$1.99; Carretta et al. 2009a) GC NGC~6397. In that case, a handful of Na-rich, Li-poor stars 
indicates a significant Li$-$Na anticorrelation (Lind et al. 2009), and a simple dilution model 
very well reproduces the Li and Na distributions (see the right-hand panel in Figure~\ref{f:6397}). 

It is important to keep in mind that TO stars in 47 Tuc are cooler 
(\teff $\sim$ 5700-5800 K) than the TO stars of NGC 6397 (\teff $\sim$ 6100-6300 K), 
and they are well below the limit usually considered for the Spite plateau. 
The scatter found in Li abundances of 47 Tuc stars is reminiscent 
(or even better, is the Population~{\sc ii} analog) of the large scatter 
of Li abundances obtained for solar twins in the old open cluster M\,67 
(Randich et al. 2000) and in general, for old {\it thick} disk stars with 
effective temperatures close to $\sim$5800 K (Ryan et al. 2001). We note, however,
that in this case the Li abundance variation is smaller with respect to typical 
values detected in Population~{\sc i} stars, probably because the 
latter ones are significantly more
metal-rich.

\section{Summary and Concluding Remarks}\label{sec:sum}

We present in this Letter Li, Na, and O abundances for a large sample ($\approx$90) of TO 
stars in the metal-rich GC 47 Tuc, providing the largest database of this 
kind available so far. Our main results can be briefly summarized as follows:
\begin{itemize}
\item[1.] We obtained a very clear Na$-$O anticorrelation, confirming the previous 
findings from giants derived by Carretta et al. (2009b). 
{\it This is the first time that the Na$-$O distributions in dwarfs and giants 
can be directly compared for a cluster using large samples of comparable sizes
for both evolutionary stages.} At least for 47 Tuc, 
evolutionary effects due to the RGB phase 
can be ruled out as contributors to the extent 
of Na$-$O anticorrelation, which in both cases span the same range 
(within the observational uncertainties).
\item[2.] As expected from stellar nucleosynthesis models 
in conjunction with multiple population scenarios, 
Li abundances should be positively correlated with O and 
anticorrelated with Na. 
At variance of the metal-poor GC NGC~6397, in the case of 47 Tuc the Li content does not show 
an anticorrelation with Na, and only a weak correlation appears with O, with 
a quite scatter distribution from both diagrams. 
Our result disagrees with the previous study by Bonifacio et al. (2007), 
who found Li$-$Na anticorrelation from a small sample of only four stars, 
and once again emphasizes the crucial role of 
statistics in this kind of analysis. 
A simple dilution model fails to reproduce the Li$-$Na$-$O distributions 
for this cluster and advocates the presence of some different mechanisms 
responsible for the observed Li pattern.
\item[3.] The scatter we find in Li abundances reminds of what has been detected, 
and reported in a large body of the literature, in Population~I stars of similar parameters 
(\teff, $\log{g}$), the most famous case being the old open cluster M~67 (see, e.g., Randich et al. 2000). 
\end{itemize}
We are not presently able to conclude if the trend we discovered in 47 Tuc
is peculiar or, on the other hand, other GCs share a similar behavior. In fact, to
date, only two GCs have been investigated from this point of view. In this context,
we mention that we cannot explain the unlikeness in the Li$-$Na 
distributions between NGC~6397 and 47 Tuc; in particular, we cannot
discriminate if the differences in \teff's or metallicity can account for such a
discrepancy. To probe this issue, it is crucial to enlarge the sample of simultaneous
determinations of Li, Na, O in GC dwarf stars, by including other (nearby) clusters
with different structural parameters (e.g., HB morphology, age, metallicity).

\begin{table*}
\begin{small}
\begin{center}
\caption{Properties of target stars.}\label{t:tab1}
\setlength{\tabcolsep}{1.3mm}
\begin{tabular}{lcccccccccccc}
\hline\hline
Star$_{\rm ID}$ &  R.A.  &  Decl.    &  $V$  &  $B$ & T$_{\rm eff}$ & $\log~g$ & $\log~n{\rm (O)}$ & Err$_{\rm O}$ & $\log~n{\rm (Na)}$ & 
Err$_{\rm {Na}}$  & $\log~n{\rm (Li)}$ & Err$_{\rm {Li}}$\\
     &  (deg) & (deg) & (mag) & (mag) & (K) & (dex) & (dex) &(dex) &(dex) &(dex) &(dex) & \\
\hline
     &        &          &        &        &      &      &        &         &        &       &       &      \\
001 & 6.15846 & $-$71.9632 & 17.347 & 17.913 & 5685 & 4.05 &  8.266 &	0.050 &  5.715 & 0.103 & 2.105 &  0.053\\	  
002 & 5.96746 & $-$71.9607 & 17.301 & 17.853 & 5676 & 4.02 &  8.061 &	0.083 &  5.741 & 0.080 & 1.538 &  0.057\\	  
003 & 6.33533 & $-$71.9429 & 17.347 & 17.913 & 5685 & 4.05 &  8.445 &	0.050 &  9.999 & 9.999 & 1.998 &  0.130\\	  
005 & 6.24487 & $-$71.9313 & 17.330 & 17.894 & 5681 & 4.04 &  8.112 &	0.111 &  5.752 & 0.080 & 2.261 &  0.096\\	  
006 & 6.16508 & $-$71.9222 & 17.372 & 17.941 & 5689 & 4.06 &  8.374 &	0.050 &  5.689 & 0.080 & 2.453 &  0.108\\	  
007 & 6.21133 & $-$71.9159 & 17.365 & 17.931 & 5688 & 4.05 &  8.493 &	0.050 &  9.999 & 9.999 & 2.381 &  0.087\\	  
008 & 6.27892 & $-$71.9146 & 17.377 & 17.940 & 5690 & 4.06 &  8.705 &	0.050 &  9.999 & 9.999 & 2.240 &  0.049\\	  
009 & 6.24204 & $-$71.9106 & 17.360 & 17.926 & 5687 & 4.05 &  8.327 &	0.050 &  9.999 & 9.999 & 2.299 &  0.081\\	  
010 & 6.11050 & $-$71.9085 & 17.335 & 17.899 & 5682 & 4.04 &  8.514 &	0.050 &  9.999 & 9.999 & 2.294 &  0.112\\	  
011 & 5.80258 & $-$71.9629 & 17.336 & 17.909 & 5682 & 4.04 &  9.999 &	9.999 &  5.910 & 0.083 & 1.987 &  0.129\\	  
\hline\hline
\end{tabular}
\end{center}
\end{small}
\tablecomments{We report our ID in Column 1, R.A. and Decl. in Columns 2 and 3, while magnitudes $V$ and $B$
are given in Columns 4 and 5. Stellar parameters (T$_{\rm eff}$ and $\log g$) along with abundances and their errors are listed 
from Columns 6 to 13, respectively. (This table is available in its entirety in a machine-readable form in the online journal. A 
portion is shown here for guidance regarding its form and content.)}
\end{table*}

\acknowledgments
We warmly thank Yazan Momany for having provided photometric data in advance of
publication. This work has been funded by PRIN INAF 2007 ``{\it Multiple Stellar Populations in Globular Clusters}" 
and DFG cluster of excellence ``Origin and Structure of the
Universe". We acknowledge the anonymous referee for her/his favorable evaluations and for providing
very constructive comments which improved the manuscript.

{}


\begin{thebibliography}{}
\bibitem[1996]{alo}Alonso, A., Arribas, S., \& Mart\'{i}nez-Roger, C. 1996, A\&AS, 117, 227
\bibitem[2003]{apl}Asplund, M., Carlsson, M., \& Botnen, A.V. 2003, A\&A, 399, 31
\bibitem[2000]{bark}Barklem, P.S., Piskunov, N., \& O'Mara, B.J. 2000, A\&A, 363, 109
\bibitem[2002]{b1}Bonifacio, P., et al. 2002, A\&A, 390, 91
\bibitem[2007]{b2}Bonifacio, P., et al. 2007, A\&A, 462, 851
\bibitem[1971]{cam}Cameron, A.G.W., \& Fowler, W.A. 1971, ApJ, 164, 111
\bibitem[2009]{car4}Carretta, E., Bragaglia, A.,  Gratton, R. , D'Orazi, V., \& Lucatello, S.  
2009a, A\&A, 508, 695
\bibitem[2004]{car1}Carretta, E., Gratton, R., Bragaglia, A., Bonifacio, P., \& Pasquini, L. 2004, A\&A, 416, 925
\bibitem[2007]{car2}Carretta, E., et al. 2007, A\&A, 464, 939
\bibitem[2009]{car3}Carretta, E., et al. 2009b, A\&A, 505, 117 
\bibitem[2010]{car5}Carretta, E., et al. 2010, ApJL, in press (arXiv:1002.1963)
\bibitem[1978]{cohe}Cohen, J.G. 1978, ApJ, 223, 487
\bibitem[2007]{dan}D'Antona, F., \& Ventura, P. 2007, MNRAS, 379, 1431
\bibitem[2007]{dec}Decressin, T., Charbonnel C., Prantzos, N., \& Ekstrom, S. 2007, A\&A,   464, 1029 
\bibitem[1990]{den}Denisenkov, P.A., \& Denisenkova, S.N. 1990, SvAL, 16, 275
\bibitem[2009]{fer}Ferraro, F.R., et al. 2009, Nature, 462, 483-486 
\bibitem[1975]{free}Freeman, K.C., \& Rodgers, A.W. 1975, ApJ, 201, 71
\bibitem[2010]{g4}Gratton, R., Carretta, E., Bragaglia, A., Lucatello, S., \& D'Orazi, V. 2010, A\&A, {\it submitted}
\bibitem[1999]{g1}Gratton, R., Carretta, E., Eriksson, K., \& Gustafsson, B. 1999, A\&A, 350, 955
\bibitem[2004]{g3}Gratton, R., Sneden, C., \& Carretta, E. 2004, ARA\&A, 42, 385
\bibitem[2001]{g2}Gratton, R., et al. 2001, A\&A, 369, 87
\bibitem[1996]{har}Harris, W.E. 1996, AJ, 112, 1487 
\bibitem[1993]{kur}Kurucz, R.L. 1993, CD-ROM 13, (Cambridge, MA: Smithsonian Astrophysical Observatory)
\bibitem[2009]{lin}Lind, K., Primas, F., Charbonnel, C., Grundahl, F., Asplund, M. 2009, 
A\&A, 503, 545 
\bibitem[2009]{mar}Marino, A.F., et al. 2009, A\&A, 505, 1099
\bibitem[2003]{mom}Momany, Y., et al. 2003, A\&A, 407, 303 
\bibitem[1997]{p1}Pasquini, L., \& Molaro, P. 1997, A\&A, 322, 109
\bibitem[2005]{p2}Pasquini, L., et al. 2005, A\&A, 441, 549
\bibitem[2006]{pra}Prantzos, N., \& Charbonnel, C. 2006, A\&A, 458, 135
\bibitem[2000]{ran}Randich, S., Pasquini, L., \& Pallavicini, R., 2000, A\&A, 356, 25			
\bibitem[2001]{rya}Ryan, S.,  Kajino, T., \& Beers, T. 2001, ApJ, 549, 55
\bibitem[2009]{vd}Ventura, P., \& D'Antona, F. 2009, A\&A, 499, 835
\end{thebibliography}
	\end{document}